\documentclass[fleqn,twoside]{article}
\usepackage{espcrc2,epsfig}

\newcommand{\sslash}[1]{#1 \hspace{-0.5em} /}
\renewcommand{\slash}[1]{#1 \hspace{-0.55em} /}
\newcommand{\Slash}[1]{#1 \hspace{-0.65em} /}

\title{Power corrections in heavy-to-light decays at large recoil energy
\thanks{Talk given at ICHEP 2002, Amsterdam, July 2002. Based on work
with A.~Chapovsky, M.~Beneke and Th.\
Feld\-mann~\protect\cite{Beneke:2002ph}}} 

\author{M. Diehl\address{Institut f\"ur Theoretische Physik E, RWTH
Aachen, 52056 Aachen, Germany} }
       
\begin{document}

\begin{abstract}
I briefly present recent work on QCD power corrections in
heavy-to-light meson decays, using an effective field theory approach.
\vspace{-0.8em}
\end{abstract}

\maketitle

\section{INTRODUCTION}

The strong interaction physics in $B$ decays is characterized by the
presence of several distance scales.  The general idea of
factorization in QCD is to separate the dynamics at short distances
(which can be treated perturbatively) from the long-distance physics
(encoded in hadronic matrix elements like decay constants or form
factors).  There are different techniques to achieve this separation.
One is the detailed analysis of Feynman graphs, used for instance in
\cite{Beneke:2000ry} to
establish QCD factorization for exclusive $B$ decays.  Here I focus on
an effective field theory approach, whose principal objects are
operators and fields.  This framework is very general and well suited
to be applied to a wide class of processes.

As an explicit application I consider exclusive $B$ decays to light
mesons (e.g.\ $B\to \pi\, \ell\nu$, $B\to K^*\, \ell^+\ell^-$) in
kinematics where the recoil energy of the light meson is large, $E\sim
m_b$.  In the limit $m_b\to \infty$ symmetry relations between the
various form factors describing the $B\to \pi$ or $B\to K^*$
transition emerge \cite{Charles:1998dr}, which greatly reduce hadronic
uncertainties when analyzing these decays in the Standard Model and
beyond.  These relations receive corrections in inverse powers of
$m_b$, which we have systematically studied in effective field theory
\cite{Beneke:2002ph}.  They also receive corrections in $\alpha_s$
from the heavy quark decay vertex (which are part of the effective
theory) and from hard interactions of the spectator quark in the $B$
meson (which have so far only been treated by direct calculation of
the relevant Feynman graphs in QCD~\cite{Beneke:2000wa}).

\section{EFFECTIVE FIELD THEORY}

\begin{figure*}[t]
\begin{center}
\psfig{width=0.7\textwidth,file=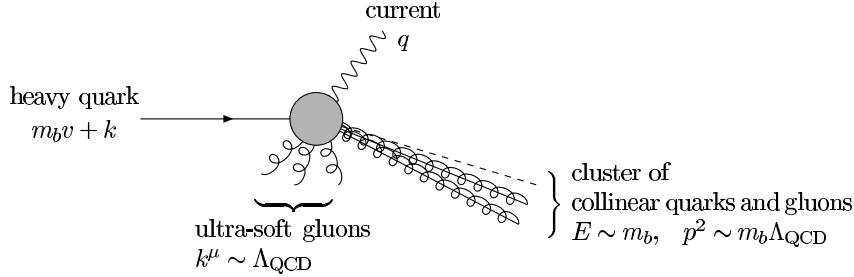}
\end{center}
\caption{\label{fig:setup}Heavy-quark decay into a cluster of
collinear and ultrasoft particles.}
\end{figure*}

To set up the effective field theory appropriate for a given process
one first identifies the relevant momentum regions in the
corresponding Feynman graphs.  One wants to keep in the theory only
the nearly on-shell momentum modes that directly connect to the
external hadrons.  In the decays of our study
(see~Fig.~\ref{fig:setup}) this is a heavy quark in the initial $B$
meson, a cluster of light quarks and gluons which move approximately
collinear to each other and form the fast outgoing meson, and
``ultrasoft'' gluons in either the initial or final state.  (At this
point we exclude the hard spectator quark interactions mentioned
above.)  Modes with large momenta $k^\mu \sim m_b$ are ``integrated
out'' from the theory: this is done in perturbation theory and
provides the Wilson coefficients associated with the shaded blob in
the figure.  The same happens to ``soft'' gluons with momenta $k^\mu
\sim \sqrt{m_b\, \Lambda_{\mathrm{QCD}}}$.  The heavy quark and its
interactions with ultrasoft gluons is described in conventional
heavy-quark effective theory (HQET), and the dynamics of collinear and
ultrasoft modes in the soft-collinear effective theory (SCET)
pioneered by Bauer et al.~\cite{Bauer:2000ew}.

For each momentum mode a separate field is introduced, e.g.\
$A_c^\mu(x)$ and $A_{\mathit{us}}^\mu(x)$ for collinear and ultrasoft
gluons.  For these fields one then must find the Lagrangian and
current operators which reproduce the results of QCD to a given
accuracy in a power-counting parameter $\lambda$, in our case given by
$\sqrt{\Lambda_{\mathrm{QCD}} /m_b}$.  It proves useful to project out
the large spinor components of fermion fields and to integrate out the
small ones.  In HQET this is the usual projection on $h_v(x) =
\frac{1}{2} (1+\sslash{v})\, Q_v(x)$, where $Q_v(x) = e^{im_b (vx)}
Q(x)$ has the large phase of the full heavy quark field removed.  For
collinear quarks the large component is
\begin{displaymath}
\xi(x) = {\textstyle\frac{1}{4}} \slash{n}_- \slash{n}_+\, q(x) ,
\end{displaymath}
where $n_-$ and $n_+$ are lightlike reference vectors, $n_-$ giving
the direction of the collinear particles and $n_+$ the opposite
direction.  In contrast to~\cite{Bauer:2000ew} we do not split off the
large phase of collinear fields in our approach~\cite{Beneke:2002ph}.
For ultrasoft quarks all spinor components are of same size and hence
kept in the theory.

To decide whether an operator is relevant to a given accuracy in
$\lambda$ one determines power counting rules for momentum components
(collinear momenta are large in the $n_-$ direction and small in the
others), for fields, their derivatives, etc.  Finally, the effective
theory ``inherits'' part of the gauge invariance of full QCD, and one
has to define appropriate gauge transformations for the effective
fields to implement this.

\subsection{The effective Lagrangian}

To obtain the effective Lagrangian for collinear and ultrasoft quarks
and gluons one can start from the QCD quark Lagrangian $\mathcal{L} =
\bar{q} i \Slash{D}\, q$, separate out the different momentum modes,
and integrate out the small spinor components of the collinear quark
field (using e.g.\ path integral methods).  The result is
\begin{displaymath}
{\cal L}_c =  \bar\xi \left( i n_- D +
  i \Slash D_\perp\, W \frac{1}{i n_+ \partial}\, W^\dag
  i\Slash D_\perp \right) \frac{\slash n_+}{2}\, \xi
\end{displaymath}
if one neglects ultrasoft quarks.  Because of the inverse derivative
$(n_+ \partial)^{-1}$ the action is nonlocal.  It is explicitly gauge
invariant: the sum $A = A_c + A_{\mathit{us}}$ of gluon fields appears
in the covariant derivative $D = \partial - ig A$ and in the Wilson
line
\begin{displaymath}
\textstyle W(x) =
P\exp\left\{ig\int_{-\infty}^0 \hspace{-0.2ex} ds \;n_+ A(x+s n_+) .
\right\}
\end{displaymath}
One must still expand $\mathcal{L}_c$ in $\lambda$: the components of
$A^\mu_c$ and $A^\mu_{\mathit{us}}$ have different scaling in
$\lambda$, which requires a Taylor expansion of $D$ and~$W$.
Furthermore, $A_{\mathit{us}}(x)$ varies more slowly in $x$ than
collinear fields with their large-momentum modes, so that
$A_{\mathit{us}}(x)$ must be appropriately Taylor expanded in $x$.

To leading order in $\lambda$ ultrasoft quarks just appear in
$\mathcal{L}_{\mathit{us}} = \bar{q}_{\mathit{us}}\,
i\Slash{D}_{\mathit{us}}\, q_{\mathit{us}}$, but in higher orders they
couple to collinear fields.  The resulting Lagrangian including
$O(\lambda)$ and $O(\lambda^2)$ correc\-tions also contains ultrasoft
Wilson lines $Z$, obtained from $W$ by setting $A_c=0$.

\subsection{The effective heavy-to-light current}

Before decaying via an electroweak current a heavy quark can radiate
collinear gluons, which puts it off shell by an amount of order
$m_b^2$.  The corresponding diagrams have no place in the effective
theory and must be summed into an effective vertex.  This leads one to
match the QCD heavy quark field onto an effective field, $Q\to e^{-i
m_b (v x)}\, Q_{\rm eff}(A_c,A_{\mathit{us}},h_v)$.  By explicitly
summing diagrams or by solving the Dirac equation for $Q_{\rm eff}$ in
an external field $A= A_c + A_{\mathit{us}}$ one obtains $Q_{\rm eff}$
as
\begin{displaymath}
W\!Z^\dag \, Q_v - \frac{1}{{\cal V}^2-1}\,
\Big( \Slash{{\cal V}} \,\Slash{{\cal V}}\; W\!Z^\dag - W\!Z^\dag \, 
   \Slash{{\cal V}}_{\rm us} \,\Slash{{\cal V}}_{\rm us} \Big) Q_v
\end{displaymath}
including corrections of order $\lambda$ and $\lambda^2$.  Here ${\cal
V}^\mu = v^\mu + i D^\mu /m_b$ is the reparametrization invariant
``velocity'' known from HQET.

Matching of the QCD light quark field onto an effective field, $q\to
q_{\rm \, eff}(A_c,A_{\mathit{us}},\xi)$, is readily performed in
SCET.  The complete QCD current is matched as $\overline{q}\, \Gamma Q
\to e^{-i m_b (v x)}\, \overline{q}{}_{\rm eff}\, \Gamma\, Q{}_{\rm
eff}$ and finally has to be Taylor expanded in $\lambda$ as in the
case of the effective Lagrangian.

\section{POWER CORRECTIONS TO FORM FACTORS}

The symmetry relations for $E, m_b\to \infty$ between the transition
form factors from $B$ to a light meson $L=\pi,\rho,K^*,\ldots$ receive
corrections in powers of $\lambda = \sqrt{\Lambda_{\mathrm{QCD}}
/m_b}$, which can be systematically described in the effective theory
of Sec.~2 by matching the QCD matrix elements $\langle L|\,
\overline{q}\, \Gamma\, Q \,| B\rangle$ onto $\langle L|\,
\overline{q}{}_{\rm \, eff} \Gamma\, Q_{\rm eff} \,| B\rangle$.
Projecting $\overline{q}_{\rm \, eff}$ and $Q_{\rm eff}$ on their
large and small spinor components one finds that
\begin{enumerate}
\item at leading order in $\lambda$ only the large components are
relevant.  This gives 1 (2) independent form factors for the
transition to a pseudoscalar (vector) meson
\cite{Charles:1998dr,Beneke:2000wa}.
\item the $O(\lambda)$ corrections involve the small components of
either $\overline{q}_{\rm \, eff}$ or $Q_{\rm eff}$, and one remains
with 2 (5) independent form factors.
\item the $O(\lambda^2)$ corrections come from the small
com\-po\-nents of both fields.  Hence no symmetry relations are left
to this accuracy, and one has 3 (7) independent form factors as in
full QCD.
\end{enumerate}
There is a caveat concerning the corrections at order $\lambda$.  The
corresponding operators in the effective theory involve different
parton configurations in the light meson than the leading-order
operators, for instance with an additional transversely polarized
gluon.  There may well be further dynamical suppression of their
matrix elements, bringing them down from $O(\lambda)$ to
$O(\lambda^2)$.  Whether this happens requires knowledge of the light
meson wave functions which goes beyond the scope of SCET.  We remark
that these putative $O(\lambda)$ corrections affect the form factor
relations used in the analysis of the forward-backward asymmetry in
$B\to K^*\, \ell^+\ell^-$, see~\cite{Burdman:1998mk}.

\section{SUMMARY}

We have performed a systematic analysis of power corrections to
soft-collinear factorization in the effective field theory formalism.
The interactions between collinear and ultrasoft quarks and gluons are
described in SCET, whose effective Lagrangian was obtained including
the power corrections of first and second order in $\lambda$.
Heavy-to-light meson decays at large recoil energy can be described by
the effective actions of SCET and HQET plus the effective
heavy-to-light current, which was also obtained including
$O(\lambda^2)$ corrections.  Aspects of the $O(\lambda)$ corrections
were previously investigated in \cite{Chay:2002vy}.

Among the 2 (5) leading-order symmetry relations between the
transition form factors for $B$ to pseudoscalar (vector) mesons, 1 (3)
receive power corrections at order $\sqrt{\Lambda_{\mathrm{QCD}}
/m_b}$.  Whether these are further suppressed dynamically remains to
be investigated.  To accuracy $\Lambda_{\mathrm{QCD}} /m_b$ no form
factor relations are left.  Phenomenological analyses of the
corresponding decays thus remain dependent on models for the relevant
hadronic physics.

\end{document}